\documentstyle[11pt]{article}
\newcommand {\be}{\begin{equation}} 
\newcommand{\fe}{\end{equation}}
\newcommand{\eqn}{\label}

\font\sm=cmr9

\def\thf{\baselineskip=\normalbaselineskip\multiply\baselineskip
by 3\divide\baselineskip by 4}

\thf

\def\spose#1{\hbox to 0pt{#1\hss}}
\def\lta{\mathrel{\spose{\lower 3pt\hbox{$\mathchar"218$}}
     \raise 2.0pt\hbox{$\mathchar"13C$}}}
\def\gta{\mathrel{\spose{\lower 3pt\hbox{$\mathchar"218$}}
    \raise 2.0pt\hbox{$\mathchar"13E$}}}

\def\ddotLambda{{\spose {\raise 2.0pt\hbox{$^{^{\bullet\bullet}}$}} 
{\Lambda}}}
\def\dotLambda{{\spose {\raise 2.0pt\hbox{$^{^{\, \bullet}}$}} {\Lambda}}}

\begin{document}

\title{FORMALLY RENORMALIZABLE GRAVITATIONALLY 
\\ SELF INTERACTING STRING MODELS}
\author{ {\bf Brandon Carter}
\\ D\'epartement d'Astrophysique Relativiste et de Cosmologie, C.N.R.S.,
\\ Observatoire de Paris, 92195 Meudon, France.}
\date{26 June 1998}

\maketitle

\begin{abstract}

It has recently been shown how the effect of the divergent part of the 
gravitational self interaction for a classical string model
in 4 dimensions can be allowed for by a renormalisation of its
stress energy tensor and in the elastic case a corresponding
renormalisation of the off shell action. It is shown here
that that it is possible to construct a new category of elastic
string models for which this effect is describable as a renormalisation
in the stricter ``formal'' sense, meaning that it only requires a rescaling
of one of the fixed parameters characterising the model. 

\end{abstract}

\section{Introduction}

The gravitational
self interaction of for a classical string model in a four dimensional
background has recently been shown~\cite{C98a} to have a divergent part that,
when suitably regularised, can be effectively be absorbed by a
renormalisation of the surface stress energy tensor of the
worldsheet.  In the particularly simple case of a Nambu-Goto
model the renormalisation is trivial in the sense that 
(contrary to what had been suggested by earlier work) the divergent part 
actually vanishes~\cite{CB98}. For more general elastic
string models, the renormalisation of the ``on-shell'' stress energy
tensor has been shown~\cite{C98b} to correspond to a non-trivial
renormalisation of the relevant variational action, in which the appropriate
adjustment has precisely the form that is obtained as the
four dimensional specialisation of a very general formula\cite{BD98}
that recently been obtained from a very different approach 
based on
an analysis of the ``off shell'' action in a background of arbitrary
dimension. 

The physical example that provided the original motivation for this line of
investigation was the case of cosmic strings described by the ``transonic''
string model~\cite{C95} that provides an effective large scale description of
the effect of short wavelength ``wiggles'' in an underlying Nambu-Goto model.
In this case the string model is qualitatively modified by the effect of the
renormalisation, in the sense that the resulting ``dressed'' model no longer
has the special transonic character (and the ensuing integrability properties)
of the original ``bare'' model.  The purpose of the present article is to
reply to the question of whether there is a category of gavitationally self
interacting elastic string models that is  renormalisable in the stricter
``formal'' sense, meaning that the algebraic form of the ``bare'' model is
qualitatively preserved in the corresponding ``dressed'' model in the sense
the the the renormalisation is describable just as a readjustment of the free
parameters specifying the particular model within the category.

\section{``Bare'' string models.}

As explained in more detail in the preceeding
references~\cite{C98b,BD98}, we are concerned with models governed by
an action ${\cal I}$ that is specified as an integral over the string
worldsheet that is specifiable as the 2-dimensional two dimensional
imbedding given by $x^\mu= \bar x^\mu\{\sigma\}$ in terms of intrinsic
coordinates $\sigma^i$ ($i=0,1$), so that the induced surface metric
will have the form $  \gamma_{ij}= g_{\mu\nu}\bar x{^\mu}{_{\! ,i}}\,
\bar x{^\nu}{_{\!,j}}$, where $g_{\mu\nu}$ is the spacetime metric of
the 4-dimensional background with local coordinates $x^\mu$. In terms
of such a worldsheet, the action integral will have the form,
 \be{\cal
I}=\int \overline{\cal L}\, \Vert\gamma\Vert^{1/2}\, d^2\sigma \,
,\eqn{2}\fe 
where $\vert\gamma\vert$ is the determinant of the induced
metric, and $\Lambda$ is the relevant Lagrangian scalar.  In the
absence of gravitational interaction -- other than what is
automatically allowed for by large scale the curvature of the
background metric $g_{\mu\nu}$ -- the Lagrangian scalar for a model of
the simple elastic kind under consideration here  would be given by a
master function $\Lambda$ depending just on the gradient of a freely
variable scalar stream function $\psi$ on the worldheet. However to
allow for the effect of shortwave gravitational perturbations
$g_{\mu\nu}\mapsto g_{\mu\nu}+\delta g_{\mu\nu}$ in terms of a
linearised gravitational field tensor $h_{\mu\nu}=\delta g_{\mu\nu}$ it
is evidently necessary to augment the Lagrangian by a corresponding
gravitational coupling term so that it takes the form 
\be\overline{\cal
L}=\Lambda +{_1\over^2} \overline T{^{\mu\nu}} h_{\mu\nu} \,
,\eqn{3}\fe
where  $T{^{\mu\nu}}$ is the relevant surface stress
energy tensor, which is given by $\overline T{^{\mu\nu}}=$
$2\Vert\gamma\Vert^{-1/2} {\partial
\big(\Lambda\Vert\gamma\Vert^{1/2}\big)/ \partial g_{\mu\nu}}$.  The
corresponding dynamical equations are given by the requirement that the
action integral should be invariant with respect to local variations of
the worldsheet imbedding and of the internal field $\psi$.

\section{``Dressed'' string models}

If the field $h_{\mu\nu}$ were due just to passing gravitational waves
from an external source the model that has just been described would
automatically be well defined and well behaved as it stands. Allowance
for self gravitation gives rise to difficulties of two kinds. The
hardest part is the evaluation of the finite long range contribution
including backreaction from emitted radiation, which, except for very
simple configurations will be tractable in practice only in a rather
approximate manner. However although it may be important in the long
run, this finite contribution will usually have a negligible effect on
the short timescale dynamics, for which it is the divergent short range
part of the self interaction that will dominate. The subject of the
present discussion is  this latter part, whose treatment requires
the introduction of a regularisation, whereby it will be 
obtained~\cite{C98a} in the form
\be\widehat h_{\mu\nu}=2\hbox{\sm G}\ \widehat l\
\big(2\overline T_{\!\mu\nu}-
\overline T_{\!\sigma}{^\sigma}g_{\mu\nu}\big)\, ,\eqn{4}\fe
where, as usual for a string self interaction in 4 dimensions, the
proportionality factor has the form $ \widehat l= {\rm ln}\big\{ {\Delta^2
/\delta_\ast^{\, 2} } \big\}$ in terms of an ``ultraviolet'' cut off
lengthscale $\delta_\ast$  representing the effective thickness of the string,
and a much larger ``infrared'' cut off $\Delta$, given by a lengthscale
characterising the large scale geometry of the string configuration.

The preceeding work~\cite{C98a} provides a system that
will be describable
in terms of the finite part $\widetilde h_{\mu\nu}=h_{\mu\nu}-
\widehat h_{\mu\nu}$ that is left over when the divergent part is 
subtracted out, by a renormalised Lagrangian of the form
\be\overline{\cal
L}=\widetilde\Lambda +{_1\over^2} \overline T{^{\mu\nu}}\widetilde
 h_{\mu\nu} \,
,\eqn{6}\fe
in which the original ``bare'' master function has been replaced by
a renormalised ``dressed'' master function expressible~\cite{C98b,BD98} by
\be \widetilde\Lambda=\Lambda+\widehat\Lambda_{\rm g}\, ,\eqn{7}\fe
in which the divergent part of the self interaction has been absorbed 
in an adjustment term of the form
\be\widehat\Lambda_{\rm g}={_1\over^4}\overline T{^{\mu\nu}}\widehat
 h_{\mu\nu} \, .\eqn{8}\fe

For any simple elastic string model of the kind considered here, the master
function $\Lambda$ will depend just on the scalar magnitude that is
specifiable~\cite{CP95} as $\chi=-\gamma^{ij}\psi_{, i}\psi_{,j}= -p_\mu
p^\mu$ where the relevant momentum vector is defined by $p^\mu=\bar
x{^\mu}_{\, , i}\gamma^{ij}\psi_{,j}$.
It can be seen that the surface stress energy tensor
of the ``bare'' model will be expressible in the form
\be \overline T{^{\mu\nu}}=\Lambda\gamma^{\mu\nu}+2{d\Lambda\over d\chi} p^\mu
p^\nu \, ,\eqn{9}\fe
using the notation $ \gamma^{\mu\nu}=\gamma^{ij}\bar x{^\mu}{_{\! ,i}}\, \bar
x{^\nu}{_{\! ,j}}$ for the fundamental tensor of the worldsheet, i.e. the
background spacetime projection of its internal metric. It can thus be seen
from (\ref{4}) that the self gravitational action contribution  (\ref{8}) will
be expressible simply as
\be \widehat\Lambda_{\rm g}=
         2\hbox{\sm G}\, \widehat l\,\Big(\chi{d\Lambda\over d\chi}\Big)^2
\eqn{11}\, .\fe

\section{Physical admissibility conditions.}

The preceeding application~\cite{C98b} of the formula (\ref{11})
was to the a smoothed
average description of the effect of shortwavelength wiggles on an underlying
Nambu-Goto string as described by the transonic model for which the relevant
Lagrangian master function will be specified by a constant mass parameter as a
function of the special form $ \Lambda=-m\sqrt{m^2-\chi}$, which
(unfortunately as far as its convenient integrability properties are
concerened) will not be preserved by the renormalisation. The purpose
of the present article is to show however that there does exist
 a simple but non-trivial category
of models whose algebraic character is preserved by the adjustment
given by (\ref{11}), and that are thus ``formally'' renormalisable
in the sense that all that is required is a rescaling of the 
constant parameters characterising the model. 
     
  The simplest example of a sub category that is ``formally''
renormalisable in this sense is of course the one constituted
by the Nambu-Goto models for which the master function is
just a constant, $\Lambda=-m^2$ where $m$  has the dimensions of
mass. There has never been any doubt about the
``formal'' renormalisability
of this subcategory, since it was always supposed that the appropriately
renormalised model would be given by another constant,
$\widetilde\Lambda=-\widetilde{m^2}$. However according to the
formula that was commonly quoted in textbooks~\cite{VS} for many
years the ``dressed'' value was given in terms of the 
the logarithmic regularisation
factor $\widehat l$ specified above by
$\widetilde{m^2}=m^2\big(1-4\hbox{\sm G}\,m^2\, \widehat l\ \big)$
whereas a more careful calculation~\cite{CB98} has recently shown
that the correct value, as obtainable directly from (\ref{11}) ,
will simply be $\widetilde{m^2}=m^2$. In other words the
renormalisation in the Nambu-Goto case is trivial in the sense that
it has no effect at all.
 
     Having made the observation that the Nambu-Goto category is --
from this point of view -- trivial, one is left with the question
of the existence of a category that would be ``formally'' renormalisable
in a non-trivial manner. As will be shown explicitly below, it is
very easy to construct a master function
 that satisfies the ``formal'' renormalisability condition, 
but what is not quite so easy is to ensure that the resulting
model also satisfies the further
requirements needed for physical admissibility. In order for the
energy densty and tension to be positive it is necessary that
the (on shell) value of the master function $\Lambda$ and of
its dual~\cite{C89a} (in the Hodge sense with respect to the 2-dimensional
geometry of the string world sheet) as given~\cite{CP95} by
\be ^\star\!\Lambda=\Lambda-2\chi{d\Lambda\over d\chi}\, ,\eqn{15}\fe
should both be negative,
\be \Lambda<0\, \hskip 1 cm ^\star\!\Lambda<0\, , \eqn{16}\fe
and in order to avoid
microscopic instability one one hand, and causality violation
on the other hand, both the extrinsic (``wiggle'' type) perturbation
propagation
speed $c_{_{\rm E}}$ and the longitudinal (sound type) perturbation
propagation speed $c_{_{\rm L}}$ must be real and less than unity
(assuming units such that the speed of light is itself unity)
the range of physical validity of an elastic string model is
restricted by the conditions~\cite{C89b}
\be 0<  c_{_{\rm E}}^{\, 2}\leq 1\, ,\hskip 1 cm
0< c_{_{\rm L}}^{\, 2}\leq 1\, ,\eqn{18}\fe
in which the quantities   $c_{_{\rm E}}^{\, 2}$
and $c_{_{\rm E}}^{\, 2}$ are given by the formulae 
\be  c_{_{\rm E}}^{\, \pm 2}={^\star\!\Lambda\over\Lambda}\, ,
\hskip 1 cm c_{_{\rm L}}^{\, \pm 2}=
- {d\,^\star\!\Lambda\over d\Lambda}\, ,\eqn{19}\fe
where the sign $\pm$ is defined to be positive, $\pm=+$, wherever
 the current
is timelike, i.e  $\chi<0$, and negative, $\pm=-$, wherever the
current is spacelike. i.e. $\chi>0$, while in the null limit
 $\chi=0$, the requirement reduces to $^\star\!\Lambda=\Lambda$
and $d\,^\star\!\Lambda /d\Lambda=-1$.

It will be convenient to simplify the foregoing formulae by a change
of variable whereby $\chi$ is replaced by a new variable
${\mit\Xi}=$ $\ln\big\{\chi/\chi_0\big\}$ for some fixed value $\chi_0$
and to use a dot to denote differentiation with respect to ${\mit \Xi}$
so that in particular one has 
\be\dotLambda=d\Lambda/d{\mit \Xi}=\chi d\Lambda/ d\chi \, .\eqn{21}\fe
This enables us to express the dual master function in the form
\be ^\star\!\Lambda=\Lambda-2\dotLambda \fe
so that one obtains
\be {^\star\!\Lambda/\Lambda}= 1-2\dotLambda/\Lambda\, ,\eqn{22}\fe
\be {d\,^\star\!\Lambda/d\Lambda}=1-2\ddotLambda/\dotLambda\, .\eqn{23}\fe
The formula for the ``dressed'' master function 
will be similarly expressible in the form
\be \widetilde\Lambda=\Lambda+
2\hbox{\sm G}\, \widehat l\,\dotLambda{^2} \, .\eqn{24} \fe
This notation can also be used to express the preceeding conditions 
(\ref{18}) for good
physical behaviour (causality and local stability)  as
\be \Lambda<2\dotLambda\leq 2\ddotLambda<\dotLambda\leq 0\, ,\eqn{25}\fe
in the timelike current regime, $\chi>0$, and as
\be \Lambda<0\leq 2\dotLambda\leq 2\ddotLambda \, ,\eqn{26}\fe
in the spacelike current regime, $\chi>0$.

\section {``Formally'' renormalisable models}

It is evident fron (\ref{24}) that, as has already been remarked, 
the action renormalisation $\Lambda\mapsto\widetilde\Lambda$ will
have no effect at all on a master function that is
constant (i.e. of Nambu-Goto type). Clearly the simplest category of functions
that will be non trivially preserved by such a transformation
consists of those that are linear in ${\mit \Xi}$, i.e. those of the form
\be \Lambda=-m^2+A{\mit\Xi}\, ,\eqn{27}\fe
where $A$ like $m^2$ is a fixed parameter. For a master function of
this type one obtains $\dotLambda=A$ and $\ddotLambda=0$, so the effect
of the action renormalisation will be expressible as a simple
parameter renormalisation $m^2\mapsto \widetilde{m^2}$,
$A\mapsto\widetilde A$, of which the latter part is trivial, 
$\widetilde A=A$ while the only non trivial part will be a mass
renormalisation given by the formula 
$\widetilde{m^2}=m^2-2\hbox{\sm G}\, \widehat l\,A^2\, .$
It can be seen however that although it thus satifies the requirement of being
renormalisable in the strict ``formal'' sense, this linear master function
does not provide a physically admissible string model, since (unless
$\dotLambda$ also vanishes) the restriction
$\ddotLambda=0$ is compatible neither with (\ref{25}) nor
(\ref{26}): as can be seen directly from (\ref{23}) it is simply inconsistent
with the requirement that the ``sound'' (longitudinal perturbation) speed
should be real.

Although the simplest mathematical possibility is thus excluded on
physical grounds, we can obtain  a ``formally renormalisable''
category of physically admissible models by going on from
linear to quadratic order in ${\mit\Xi}$. There is no loss of generality
in writing a quadratic function of ${\mit\Xi}$ in the form
\be \Lambda=-m^2+C{\mit\Xi}^2\, ,\eqn{29}\fe
where $C$ like $m^2$ is a fixed parameter. (The reason why, provided the
coefficient $C$ is non-zero, no generality can be gained by adding in an extra
linear term $A{\mit\Xi}$, is that such a term  could be absorbed into the
homogeneously quadratic part by a rescaling of the constant parameter $\chi_0$
that was used to fix the calibration of ${\mit\Xi}$).  For a master function
of this quadratic type one obtains $\dotLambda=2C{\mit\Xi}$ and
$\ddotLambda=2C$. The effect of the action renormalisation will therefore be
expressible as a simple parameter renormalisation $m^2\mapsto
\widetilde{m^2}$, $C\mapsto\widetilde C$, of which in this case it is the
first part that is trivial, $\widetilde{m^2}=m^2$, while the second part will
have the non-trivial form 
\be\widetilde C= C+8\hbox{\sm G}\, \widehat l\, C^2\, .\fe

In order for a master function of the form (\ref{29}) to provide
a string model satisfying the physical admissibility conditions
recpitulated above, it can be seen to be necessary and sufficient
that the parameter $m$ should be non zero,
and that the other parameter $C$ should satisfies the condition
\be 3C>-m^2\, .\fe
When $C$ is negative, $\chi$ will also have to be negative, i.e. the
current is restricted to be timelike, and the permissible
range for ${\mit\Xi}$ will be given by
\be -{m^2\over 3}<C<-{m^2\over 4} \hskip 1 cm \Rightarrow\hskip 1 cm
1\leq {\mit \Xi}< 2\left(1-\sqrt{1-{m^2\over 4\vert C\vert}}\right) \, ,\fe
(the lower limit being where $c_{_{\rm L}}\rightarrow 1$ and the upper 
limit  where $c_{_{\rm E}}\rightarrow 0$) and
\be -{m^2\over 4} \leq C<0 \hskip 1 cm \Rightarrow \hskip 1 cm
1\leq {\mit\Xi}<2\, ,\fe
(the lower limit being again where $c_{_{\rm L}}\rightarrow 1$
and the upper limit where $c_{_{\rm L}}\rightarrow 0$).
When $C$ is positive, $\chi$ will also have to be positive, i.e. the
current is restricted to be spacelike, and the permissible range for
${\mit\Xi}$ will be given by
\be 0 <C\leq m^2 \hskip 1 cm \Rightarrow\hskip 1 cm
0\leq {\mit\Xi}\leq 1\, ,\fe
(the lower limit being where $c_{_{\rm E}}\rightarrow 1$
and the upper limit where $c_{_{\rm L}}\rightarrow 1$)
and
\be C>m^2 \hskip 1 cm \Rightarrow\hskip 1 cm
          0\leq {\mit\Xi}<{m\over\sqrt C}\, ,\fe
(the lower limit being again where $c_{_{\rm E}}\rightarrow 1$
and the upper limit where $c_{_{\rm E}}\rightarrow 0$).

The quadratic formula (\ref{29}) (for $C>-m^2/3$) provides not only
the lowest order function of ${\mit\Xi}$ that gives a ``formally''
renormable string model that is physically admissible, but also
the highest order function with this property: for example
if we were to include a cubic order term in the ``bare'' master function,
the corresponding ``dressed'' master function would be of
not cubic but quartic order.

I would like to thank Richard Battye, Thibault Damour, Patrick Peter,
and Gary Gibbons for stimulating conversations.

\vfill\eject

\end{document}